\documentclass{PoS}
\usepackage{pdfpages}
\usepackage{booktabs}

\title{First combined search for neutrino point-sources in the Southern Sky with the ANTARES and IceCube neutrino telescopes}

\ShortTitle{ANTARES-IceCube combined point-source search}

\author{\Large The ANTARES$^{1}$ and IceCube$^{2}$ Collaborations\\ \scriptsize{$^{1}$ Complete list of authors on {pages 2-4}   } \\ \scriptsize{$^{2}$ Complete list of authors on pages 5-7} }

\abstract{A search for cosmic neutrino point-like sources using the ANTARES and IceCube neutrino telescopes over the Southern Hemisphere is presented. The ANTARES data was collected between January 2007 and December 2012, whereas the IceCube data ranges from April 2008 to May 2011. Clusters of muon neutrinos over the diffusely distributed background have been looked for by means of an unbinned maximum likelihood maximisation. This method is used to search for a localised excess of events over the whole Southern Sky assuming an $E^{-2}$ source spectrum. A search over a pre-selected list of candidate sources has also been carried out for different source assumptions: spectral indices of 2.0 and 2.5, and energy cutoffs of 1 PeV, 300 TeV and 100 TeV. No significant excess over the expected background has been found, and upper limits for the candidate sources are presented compared to the individual experiments.

\vspace{4mm}
{\bf Corresponding authors:}
\speaker{Javier Barrios-Mart\'i $^{1a}$},
Chad Finley$^{2b}$\\
{\it
$^1$ javier.barrios@ific.uv.es \\
$^2$ cfinley@fysik.su.se \\
} \\
     \llap{$^a$}Instituto de F\'isica Corpuscular, IFIC (UV-CSIC), Parque Cient\'ifico, C/Catedr\'atico Jos\'e Beltr\'an 2, E-46980 Paterna, Spain \\
     \llap{$^b$}Oskar Klein Centre and Dept. of Physics, Stockholm University, SE-10691 Stockholm, Sweden
}

\FullConference{The 34th International Cosmic Ray Conference,\\
		30 July- 6 August, 2015\\
		The Hague, The Netherlands}

\begin{document}

\newcommand\invisiblesection[1]{%
  \refstepcounter{section}%
%  \addcontentsline{toc}{section}{\protect\numberline{\thesection}#1}%
  \cftaddtitleline{toc}{section}{\protect\numberline{\thesection}#1}{\thepage}%
  \sectionmark{#1}}

\newcommand\invisiblesectionnopagenumber[2]{%
  \refstepcounter{section}%
  \cftaddtitleline{toc}{section}{\protect\numberline{\thesection}#1}{#2}%
  \sectionmark{#1}}

\noindent{\bf\LARGE The ANTARES Collaboration}\\[3mm]
S.~Adri\'an-Mart\'inez$^a$,
M.~Ageron$^g$,
A.~Albert$^b$, 
M.~Andr\'e$^c$, 
%M.~Anghinolfi$^d$,
%I. Al Samarai$^g$,
G.~Anton$^e$, 
M.~Ardid$^a$, 
%T.~Astraatmadja$^f$,
% \author[Erlangen]{L.~Anton}
J.-J.~Aubert$^g$,
B.~Baret$^h$,
J.~Barrios-Mart\'{\i}$^i$,
S.~Basa$^j$,
V.~Bertin$^g$,
S.~Biagi$^{k,l}$,
R.~Bormuth$^{g,ak}$,
%C.~Bigongiari$^i$,
%C.~Bogazzi$^f$,
%B.~Bouhou$^h$,
M.C.~Bouwhuis$^f$,
R. Bruijn$^{f,ac}$,
J.~Brunner$^g$,
J.~Busto$^g$,
A.~Capone$^{m,n}$,
L.~Caramete$^o$,
%C.~C$\mathrm{\hat{a}}$rloganu$^p$,
J.~Carr$^g$,
%Ph.~Charvis$^q$,
T.~Chiarusi$^k$,
M.~Circella$^r$,
%F.~Classen$^e$,
%L.~Core$^g$,
A.~Coleiro$^h$,
R.~Coniglione$^v$,
H.~Costantini$^g$,
P.~Coyle$^g$,
A.~Creusot$^h$,
%C.~Curtil$^g$,
I.~Dekeyser$^s$,
%G.~Derosa$^{t,u}$,
A.~Deschamps$^q$,
G.~De~Bonis$^{m,n}$,
{C.~Distefano$^v$,
C.~Donzaud$^{h,w}$,
D.~Dornic$^g$,
%Q.~Dorosti$^x$,
D.~Drouhin$^b$,
A.~Dumas$^p$,
T.~Eberl$^e$,
D.~Els\"asser$^y$,
%U.~Emanuele$^i$,
A.~Enzenh\"ofer$^e$,
%J.-P.~Ernenwein$^g$,
S%.~Escoffier$^g$,
K.~Fehn$^e$,
I.~Felis$^a$,
P.~Fermani$^{m,n}$,
%F.~Folger$^e$,
L.A.~Fusco$^{k,l}$,
S.~Galat\`a$^h$,
P.~Gay$^p$,
S.~Gei{\ss}els\"oder$^e$,
K.~Geyer$^e$,
V.~Giordano$^z$,
A.~Gleixner$^e$,
H.~Glotin$^{an}$,
R.~Gracia-Ruiz$^h$,
%J.P.~ G\'omez-Gonz\'alez$^i$,
K.~Graf$^e$,
%G.~Guillard$^p$,
S.~Hallmann$^e$,
H.~van~Haren$^{aa}$,
A.J.~Heijboer$^f$,
Y.~Hello$^q$,
J.J.~Hern\'andez-Rey$^i$,
%B.~Herold$^e$,
J.~H\"o{\ss}l$^e$,
J.~Hofest\"adt$^e$,
C.~Hugon$^d$,
C.W~James$^e$,
M.~de~Jong$^f$,
M.~Kadler$^y$,
O.~Kalekin$^e$,
%A.~Kappes$^e$,
U.~Katz$^e$,
D.~Kie{\ss}ling$^e$,
P.~Kooijman$^{f,ab,ac}$,
A.~Kouchner$^h$,
M.~Kreter$^y$,
I.~Kreykenbohm$^{ad}$,
V.~Kulikovskiy$^{d,ae}$,
C.~Lachaud$^h$,
%R.~Lahmann$^e$,
%E.~Lambard$^g$,
%G.~Lambard$^i$,
%G.~Larosa$^a$,
%D.~Lattuada$^v$,
D. ~Lef\`evre$^s$,
E.~Leonora$^{z,af}$,
%H.~Loehner$^x$,
S.~Loucatos$^{ah}$,
%S.~Mangano$^i$,
M.~Marcelin$^j$,
A.~Margiotta$^{k,l}$,
A.~Marinelli$^{ao,ap}$,
J.A.~Mart\'inez-Mora$^a$,
%S.~Martini$^s$,
A.~Mathieu$^g$,
T.~Michael$^f$,
P.~Migliozzi$^t$,
A.~Moussa$^{am}$,
L.~Moscoso$^{h,\dagger}$,
C.~Mueller$^y$,
%H.~Motz$^e$,
%C.~Mueller$^{ad,y}$,
%M.~Neff$^e$,
E.~Nezri$^j$,
%D.~Palioselitis$^f$,
G.E.~P\u{a}v\u{a}la\c{s}$^o$,
P.~Payre$^{g,\dagger}$,
C.~Pellegrino$^{k,l}$,
C.~Perrina$^{m,n}$,
P.~Piattelli$^v$,
V.~Popa$^o$,
T.~Pradier$^{ai}$,
C.~Racca$^{b}$,
G.~Riccobene$^v$,
K.~Roensch$^e$,
%R.~Richter$^e$,
%C.~Rivi\`ere$^g$,
%K.~Roensch$^e$,
%A.~Rostovtsev$^{aj}$,
M.~Salda\~{n}a$^a$,
D.F.E.~Samtleben$^{f,ak}$,
A.~S{\'a}nchez-Losa$^i$,
M.~Sanguineti$^{d,al}$,
P.~Sapienza$^v$,
J.~Schmid$^e$,
J.~Schnabel$^e$,
%S.~Schulte$^f$,
F.~Sch\"ussler$^{ah}$,
T.~Seitz$^e$,
%R.~Shanidze$^e$,
C.~Sieger$^e$,
%A.~Spies$^e$,
M.~Spurio$^{k,l}$,
J.J.M.~Steijger$^f$,
Th.~Stolarczyk$^{ah}$,
%D.~Stransky$^e$,
M.~Taiuti$^{d,al}$,
C.~Tamburini$^s$,
%Y.~Tayalati$^{am}$,
A.~Trovato$^v$,
M.~Tselengidou$^e$,
D.~Turpin$^g$,
C.~T\"onnis$^i$,
B.~Vallage$^{ah}$,
C.~Vall\'ee$^g$,
V.~Van~Elewyck$^h$,
E.~Visser$^f$,
D.~Vivolo$^{t,u}$,
S.~Wagner$^e$,
J.~Wilms$^{ad}$,
%E.~de~Wolf$^{f,ac}$,
%K.~Yatkin$^g$,
%H.~Yepes$^i$,
J.D.~Zornoza$^i$,
J.~Z\'u\~{n}iga$^i$}

% \fntext[tag:1]{\scriptsize{Also at University of Leiden, the Netherlands}}
% \fntext[tag:2]{\scriptsize{On leave of absence at the Humboldt-Universit\"at zu Berlin}}
% \fntext[tag:3]{\scriptsize{Also at Accademia Navale de Livorno, Livorno, Italy}}

%UPV
\vspace{5mm}
%All the affiliations.
\noindent
$^a$ Institut d'Investigaci\'o per a la Gesti\'o Integrada de les Zones Costaneres (IGIC) - Universitat Polit\`ecnica de Val\`encia. C/  Paranimf 1 , 46730 Gandia, Spain\\
%Colmar
$^b$ GRPHE -Universit\'e de Haute Alsace \& Institut universitaire de technologie de Colmar, 34 rue du Grillenbreit BP 50568 - 68008 Colmar, France\\
%UPC
$^c$ Technical University of Catalonia, Laboratory of Applied Bioacoustics, Rambla Exposici\'o,08800 Vilanova i la Geltr\'u,Barcelona, Spain\\
%Genova
$^d$ INFN - Sezione di Genova, Via Dodecaneso 33, 16146 Genova, Italy\\
%Erlangen
$^e$Friedrich-Alexander-Universit\"at Erlangen-N\"urnberg, Erlangen Centre for Astroparticle Physics, Erwin-Rommel-Str. 1, 91058 Erlangen, Germany\\
%NIKHEF
$^f$Nikhef, Science Park,  Amsterdam, The Netherlands\\
%CPPM
$^g$Aix Marseille Universit\'e, CNRS/IN2P3, CPPM UMR 7346, 13288, Marseille, France\\
%APC
$^h$APC, Universit\'e Paris Diderot, CNRS/IN2P3, CEA/IRFU, Observatoire de Paris, Sorbonne Paris Cit\'e, 75205 Paris, France\\
%IFIC
$^i$IFIC - Instituto de F\'isica Corpuscular, Parque Cient\'ifico
c/ Catedr\'atico Jos\' e Beltr\'an, 2 - E46980 Paterna, Valencia (Spain)\\
%LAM
$^j$LAM - Laboratoire d'Astrophysique de Marseille, P\^ole de l'\'Etoile Site de Ch\^ateau-Gombert, rue Fr\'ed\'eric Joliot-Curie 38,  13388 Marseille Cedex 13, France\\
%Bologna
$^k$INFN - Sezione di Bologna, Viale Berti-Pichat 6/2, 40127 Bologna, Italy\\
%Bologna-UNI
$l$Dipartimento di Fisica dell'Universit\`a, Viale Berti Pichat 6/2, 40127 Bologna, Italy\\
%roma
$^m$INFN -Sezione di Roma, P.le Aldo Moro 2, 00185 Roma, Italy\\
%Roma UNI
$^n$Dipartimento di Fisica dell'Universit\`a La Sapienza, P.le Aldo Moro 2, 00185 Roma, Italy\\
%ISS
$^o$Institute for Space Sciences, R-77125 Bucharest, M\u{a}gurele, Romania\\
%Clermont
$^p$Laboratoire de Physique Corpusculaire, Clermont Universit\'e, Universit\'e Blaise Pascal, CNRS/IN2P3, BP 10448, F-63000 Clermont-Ferrand, France\\
%GEOAZUR
$^q$G\'eoazur, Universit\'e Nice Sophia-Antipolis, CNRS, IRD, Observatoire de la C\^ote d'Azur, Sophia Antipolis, France \\
%Bari
$^r$INFN - Sezione di Bari, Via E. Orabona 4, 70126 Bari, Italy\\
%COM
$^s$Aix Marseille Universit\'e, CNRS/INSU, IRD, Mediterranean Institute of Oceanography (MIO), UM 110, Marseille, France ; Universit\'e de Toulon, CNRS, IRD, Mediterranean Institute of Oceanography (MIO), UM 110, La Garde, France\\
%Napoli
$^t$INFN -Sezione di Napoli, Via Cintia 80126 Napoli, Italy\\
%Napoli UNI
$^u$Dipartimento di Fisica dell'Universit\`a Federico II di Napoli, Via Cintia 80126, Napoli, Italy\\
%LNS
$^v$INFN - Laboratori Nazionali del Sud (LNS), Via S. Sofia 62, 95123 Catania, Italy\\
%UPS
$^w$Univ. Paris-Sud , 91405 Orsay Cedex, France\\
%KVI
%$^x$Kernfysisch Versneller Instituut (KVI), University of Groningen, Zernikelaan 25, 9747 AA Groningen, The Netherlands\\
%Wuerzburg
$^y$Institut f\"ur Theoretische Physik und Astrophysik, Universit\"at W\"urzburg, Emil-Fischer Str. 31, 97074 W\"urzburg, Germany\\
%Catania
$^z$INFN - Sezione di Catania, Viale Andrea Doria 6, 95125 Catania, Italy\\
%NIOZ
$^{aa}$Royal Netherlands Institute for Sea Research (NIOZ), Landsdiep 4,1797 SZ 't Horntje (Texel), The Netherlands\\
%UU
$^{ab}$Universiteit Utrecht, Faculteit Betawetenschappen, Princetonplein 5, 3584 CC Utrecht, The Netherlands\\
%UvA
$^{ac}$Universiteit van Amsterdam, Instituut voor Hoge-Energie Fysica, Science Park 105, 1098 XG Amsterdam, The Netherlands\\
%Bamberg
$^{ad}$Dr. Remeis-Sternwarte and ECAP, Universit\"at Erlangen-N\"urnberg,  Sternwartstr. 7, 96049 Bamberg, Germany\\
%MSU
$^{ae}$Moscow State University,Skobeltsyn Institute of Nuclear Physics,Leninskie gory, 119991 Moscow, Russia\\
%Catania UNI
$^{af}$Dipartimento di Fisica ed Astronomia dell'Universit\`a, Viale Andrea Doria 6, 95125 Catania, Italy\\
%IRFU/SPP
$^{ah}$Direction des Sciences de la Mati\`ere - Institut de recherche sur les lois fondamentales de l'Univers - Service de Physique des Particules, CEA Saclay, 91191 Gif-sur-Yvette Cedex, France\\
%IPHC
$^{ai}$Universit\'e de Strasbourg, IPHC, 23 rue Becquerel 67087 Strasbourg, France
CNRS, UMR7178, 67087 Strasbourg, France\\
%ITEP
%$^{aj}$ITEP - Institute for Theoretical and Experimental Physics, B. Cheremushkinskaya 25, 117218 Moscow, Russia\\
%Leiden
$^{ak}$Universiteit Leiden, Leids Instituut voor Onderzoek in Natuurkunde, 2333 CA Leiden, The Netherlands\\
%Genova-UNI
$^{al}$Dipartimento di Fisica dell'Universit\`a, Via Dodecaneso 33, 16146 Genova, Italy\\
%LPMR
$^{am}$University Mohammed I, Laboratory of Physics of Matter and Radiations, B.P.717, Oujda 6000, Morocco\\
$^{an}$
LSIS, Aix Marseille Universit\'e CNRS ENSAM LSIS UMR 7296 13397 Marseille, France ; Universit\'e de Toulon CNRS LSIS UMR 7296 83957 La Garde, France ; Institut Universitaire de France, 75005 Paris, France\\
$^{ao}$INFN - Sezione di Pisa, Largo B. Pontecorvo 3, 56127 Pisa, Italy\\
$^{ap}$Dipartimento di Fisica dell'Universit\`a, Largo B. Pontecorvo 3, 56127 Pisa, Italy\\

$^{\dagger}$Deceased\\
%\end{frontmatter}
% \end{document}

\vspace{5mm}
\noindent{\bf Acknowledgment:}
The authors acknowledge the financial support of the funding agencies:
Centre National de la Recherche Scientifique (CNRS), Commissariat \`a
l'\'ener\-gie atomique et aux \'energies alternatives (CEA),
Commission Europ\'eenne (FEDER fund and Marie Curie Program), R\'egion
\^Ile-de-France (DIM-ACAV) R\'egion Alsace (contrat CPER), R\'egion
Provence-Alpes-C\^ote d'Azur, D\'e\-par\-tement du Var and Ville de La
Seyne-sur-Mer, France; Bundesministerium f\"ur Bildung und Forschung
(BMBF), Germany; Istituto Nazionale di Fisica Nucleare (INFN), Italy;
Stichting voor Fundamenteel Onderzoek der Materie (FOM), Nederlandse
organisatie voor Wetenschappelijk Onderzoek (NWO), the Netherlands;
Council of the President of the Russian Federation for young
scientists and leading scientific schools supporting grants, Russia;
National Authority for Scientific Research (ANCS), Romania;
Mi\-nis\-te\-rio de Econom\'{\i}a y Competitividad (MINECO), Prometeo
and Grisol\'{\i}a programs of Generalitat Valenciana and MultiDark,
Spain; Agence de  l'Oriental and CNRST, Morocco. We also acknowledge
the technical support of Ifremer, AIM and Foselev Marine for the sea
operation and the CC-IN2P3 for the computing facilities
\clearpage
\noindent{\bf\LARGE IceCube Collaboration}\\[3mm]
M.~G.~Aartsen$^{2}$,
K.~Abraham$^{32}$,
M.~Ackermann$^{48}$,
J.~Adams$^{15}$,
J.~A.~Aguilar$^{12}$,
M.~Ahlers$^{29}$,
M.~Ahrens$^{39}$,
D.~Altmann$^{23}$,
T.~Anderson$^{45}$,
M.~Archinger$^{30}$
% ...

\vspace{5mm}
%All the affiliations.
\noindent
$^{1}$III. Physikalisches Institut, RWTH Aachen University, D-52056 Aachen, Germany\\
$^{2}$School of Chemistry \& Physics, University of Adelaide, Adelaide SA, 5005 Australia\\
% ...
$^{12}$Universit\'e Libre de Bruxelles, Science Faculty CP230, B-1050 Brussels, Belgium\\
% ...
$^{15}$Dept.~of Physics and Astronomy, University of Canterbury, Private Bag 4800, Christchurch, New Zealand\\
% ...
$^{23}$Erlangen Centre for Astroparticle Physics, Friedrich-Alexander-Universit\"at Erlangen-N\"urnberg, D-91058 Erlangen, Germany\\
% ...
$^{29}$Dept.~of Physics and Wisconsin IceCube Particle Astrophysics Center, University of Wisconsin, Madison, WI 53706, USA\\
$^{30}$Institute of Physics, University of Mainz, Staudinger Weg 7, D-55099 Mainz, Germany\\
% ...
$^{32}$Technische Universit\"at M\"unchen, D-85748 Garching, Germany\\
% ...
$^{39}$Oskar Klein Centre and Dept.~of Physics, Stockholm University, SE-10691 Stockholm, Sweden\\
% ...
$^{45}$Dept.~of Physics, Pennsylvania State University, University Park, PA 16802, USA\\
% ...
$^{48}$DESY, D-15735 Zeuthen, Germany\\
{\scriptsize
$^{a}$Earthquake Research Institute, University of Tokyo, Bunkyo, Tokyo 113-0032, Japan\\
$^{b}$NASA Goddard Space Flight Center, Greenbelt, MD 20771, USA\\
}

\vspace{5mm}
\noindent{\bf Acknowledgment:}
We acknowledge the support from the following agencies:
U.S. National Science Foundation-Office of Polar Programs,
U.S. National Science Foundation-Physics Division,
University of Wisconsin Alumni Research Foundation,
the Grid Laboratory Of Wisconsin (GLOW) grid infrastructure at the University of Wisconsin - Madison, the Open Science Grid (OSG) grid infrastructure;
U.S. Department of Energy, and National Energy Research Scientific Computing Center,
the Louisiana Optical Network Initiative (LONI) grid computing resources;
Natural Sciences and Engineering Research Council of Canada,
WestGrid and Compute/Calcul Canada;
Swedish Research Council,
Swedish Polar Research Secretariat,
Swedish National Infrastructure for Computing (SNIC),
and Knut and Alice Wallenberg Foundation, Sweden;
German Ministry for Education and Research (BMBF),
Deutsche Forschungsgemeinschaft (DFG),
Helmholtz Alliance for Astroparticle Physics (HAP),
Research Department of Plasmas with Complex Interactions (Bochum), Germany;
Fund for Scientific Research (FNRS-FWO),
FWO Odysseus programme,
Flanders Institute to encourage scientific and technological research in industry (IWT),
Belgian Federal Science Policy Office (Belspo);
University of Oxford, United Kingdom;
Marsden Fund, New Zealand;
Australian Research Council;
Japan Society for Promotion of Science (JSPS);
the Swiss National Science Foundation (SNSF), Switzerland;
National Research Foundation of Korea (NRF);
Danish National Research Foundation, Denmark (DNRF)

\clearpage

%\invisiblesection{Search for High Energy Neutron Point Sources with IceTop --- PoS(ICRC2015)0250}
\section{Introduction}

Neutrinos offer unique insight into the Universe due to the fact that they interact only weakly. This also implies that their detection is challenging. The field is presently led by the IceCube \cite{icecube} and ANTARES \cite{AntDetect} experiments. IceCube is the first detector to reach the cubic-kilometer size predicted to be necessary to detect cosmic neutrino fluxes. Recently, IceCube has reported the crucial discovery of a flux of neutrinos  up to $\sim$\,PeV energies which cannot be explained by the background of atmospheric muons and neutrinos \cite{%IceCube1
IceCube_HESE_2yr, IceCube_HESE_3yr}. Meanwhile the ANTARES experiment has proven the feasibility of the Cherenkov telescope technique in sea water \cite{ANTARES-OSCI, ANTARES-MUAT}.
While its instrumented volume is significantly smaller than that of IceCube, its geographical location provides a better view of the Southern sky for neutrino energies below 100 TeV. This provides better sensitivity to the many predicted Galactic sources of neutrinos in this part of the sky.  The complementarity of the detectors for Southern sky sources allows for a gain in sensitivity by combining the analysis of data from both experiments in a joint search for point sources.
The improvement with this combination depends on the actual details of the fluxes, in particular the energy spectrum and a possible energy cut-off of the signal. The energy spectra are not yet known and predictions vary widely depending on the source model.

\section{Neutrino Data Samples}\label{sample}

The data sample corresponds to all events from the Southern sky which were included in the three-year IceCube point-source analysis \cite{PS-IceCube-79} combined with the events in the latest ANTARES point-source analysis \cite{PS-ANTARES}. 
 The ANTARES sample corresponds to data recorded from 2007 January 29 to 2012 December 31. The total number of events in this sample amounts to 5516, of which 4136 are from the Southern Hemisphere. The estimated contamination of mis-reconstructed atmospheric muons is of 10\%. 
 The IceCube data was recorded from 2008 April 5 to 2011 May 13, with a total number of 146\,018 events in the Southern Sky. In contrast to the ANTARES sample, these events are predominantly atmospheric muons rather than atmospheric neutrinos, because the Earth cannot be used as a neutrino filter for directions above the detector.

The fraction of expected source events needs to be calculated in order to estimate the relative contribution of each sample in the likelihood. This quantity is defined as the ratio of the expected number of signal events from the given sample to the expected number from all samples,

\begin{equation}
C^j (\delta, {d\Phi}/{dE_\nu} ) = \frac{N^j(\delta, {d\Phi}/{dE_\nu})}{\sum_{i} N^i(\delta, {d\Phi}/{dE_\nu})} ,
\end{equation}

\noindent where the total number of expected events for the $j$-th sample, $N^j$, with a given source declination, $\delta$, and a given source spectrum, $\frac{d\Phi}{dE_\nu}$, can be calculated as

\begin{equation}\label{eq:acc}
N^j \left(\delta, \frac{d\Phi}{dE_\nu} \right) = \int dt \int dE_\nu A^{j}_{\rm eff}(E_\nu,\delta) \frac{d\Phi}{dE_\nu} \; .
\end{equation}

The time integration extends over the live time of each sample and $A^{j}_{\mathrm{eff}}(E_\nu, \delta)$ indicates the effective area of the corresponding detector layout $j$ as a function of the neutrino energy, $E_\nu$, and the declination of the source, $\delta$. The declination of a given event is not directly related with the zenith direction in the ANTARES telescope, and therefore, the effective area for a given declination changes at different times of the day. Steady, non time-dependent sources are assumed for this analysis. Therefore, it is possible to integrate the zenith dependence for the considered period.

\begin{figure}[!h]
	\centering
	\includegraphics[width=.4\textwidth]{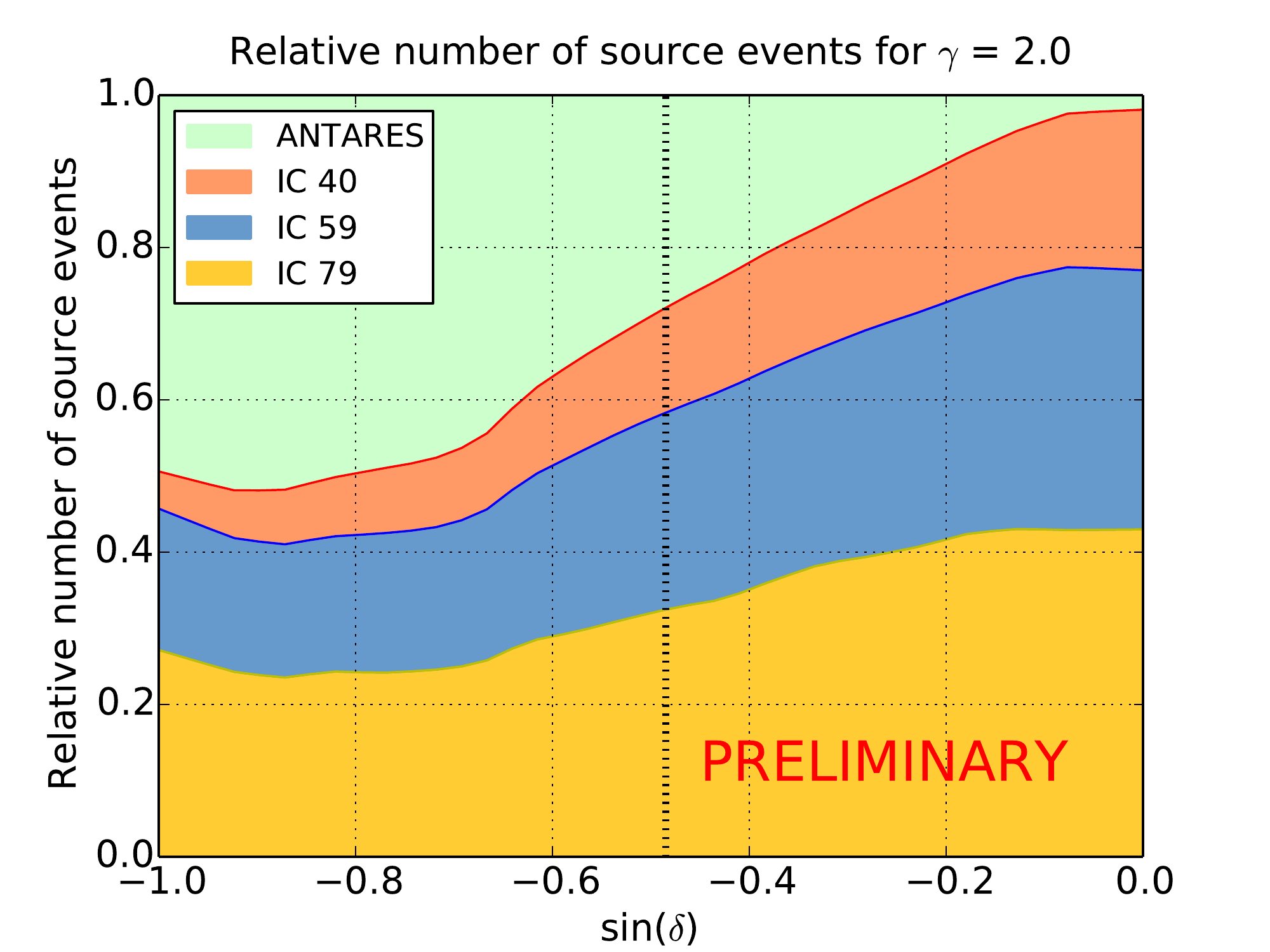}
	\caption{Relative fraction of signal events for each sample as a function of the source declination for the case of an $E^{-2}$ energy spectrum. The orange, blue, and yellow shaded areas correspond respectively to the IceCube 40, 59 and 79-string data samples, and the green shaded area indicates the ANTARES sample. The relative fraction of signal events is used as part of the likelihood function calculation during the search.}
	\label{RelContX20}
	\vspace{-11pt}
\end{figure}

Since each detector layout has a different response depending on the neutrino energy and declination, this relative fraction of source events needs to be calculated for different source spectra and source declinations. Figure \ref{RelContX20} shows the relative fraction of signal events for an unbroken $E^{-2}$ spectrum, which corresponds to the standard first order Fermi spectrum \cite{Fermi-1, Fermi-2}. In this case, there is a significant contribution from all samples over most of the Southern Sky, with the ANTARES contribution being more significant for declinations closer to $\delta$ = --90$^\circ$, and IceCube for declinations closer to 0$^\circ$. 

\begin{figure}[!h]
   \centering
	\includegraphics[width=.35\textwidth]{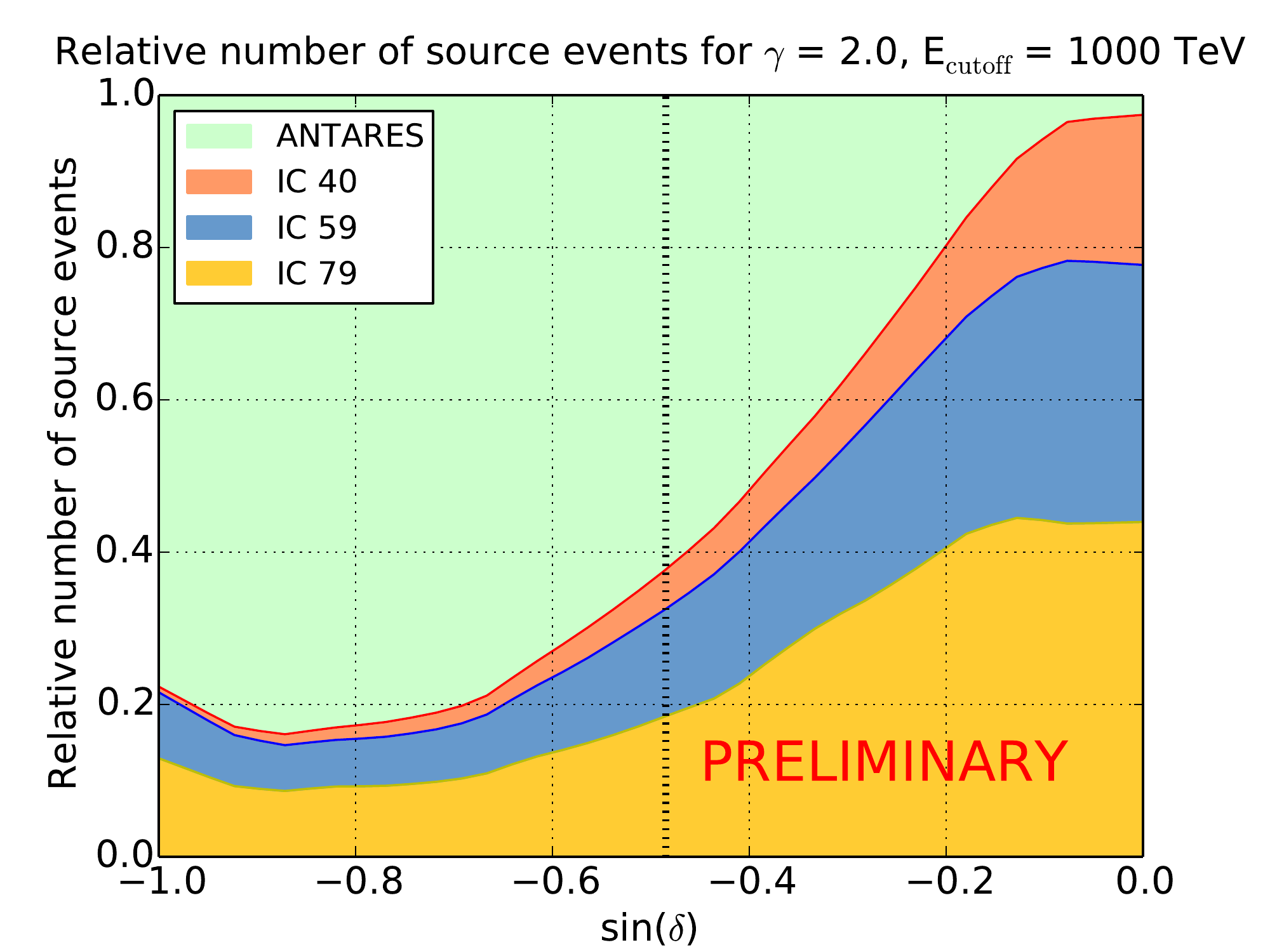}
	\includegraphics[width=.35\textwidth]{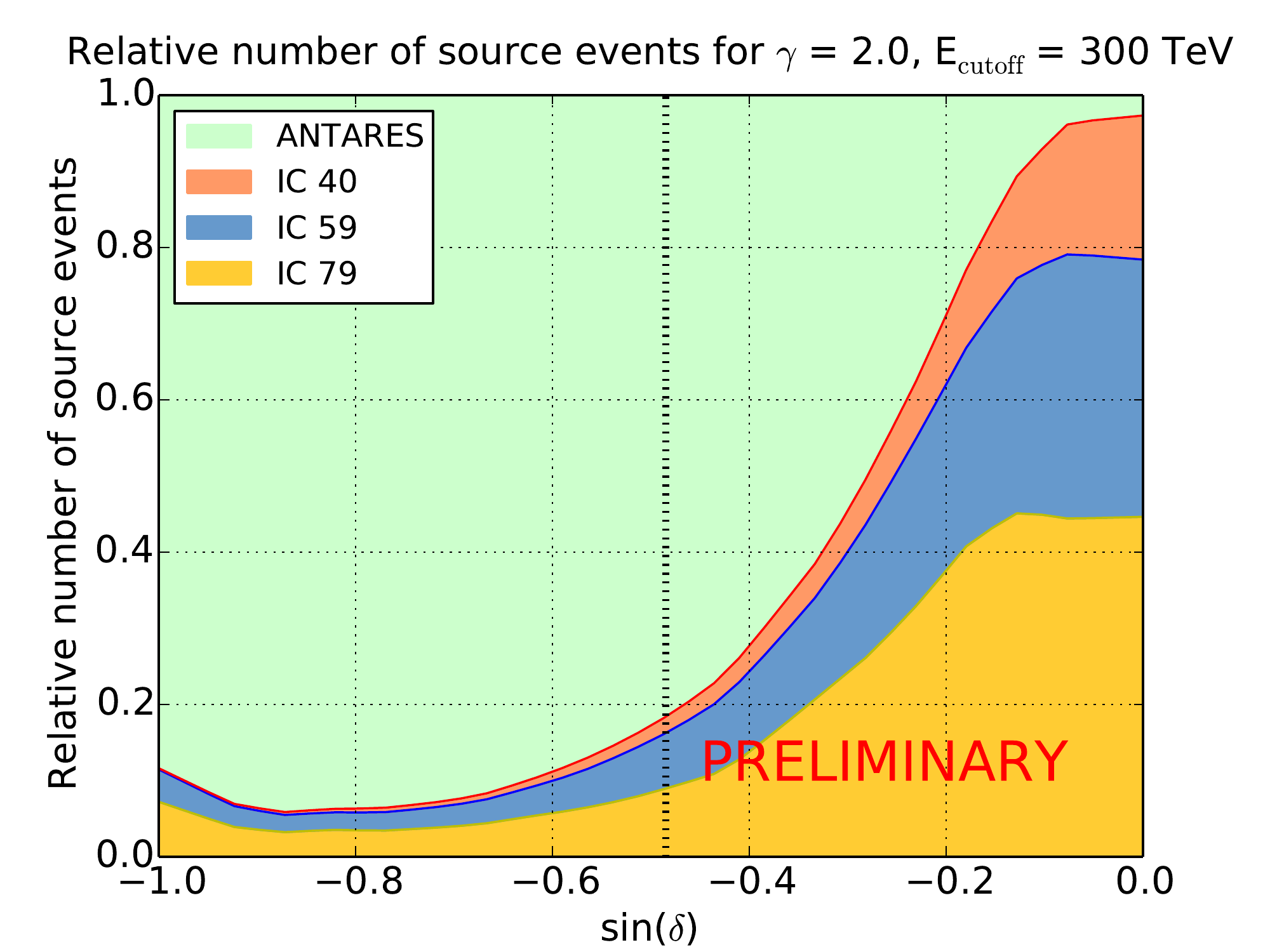}
    \includegraphics[width=.35\textwidth]{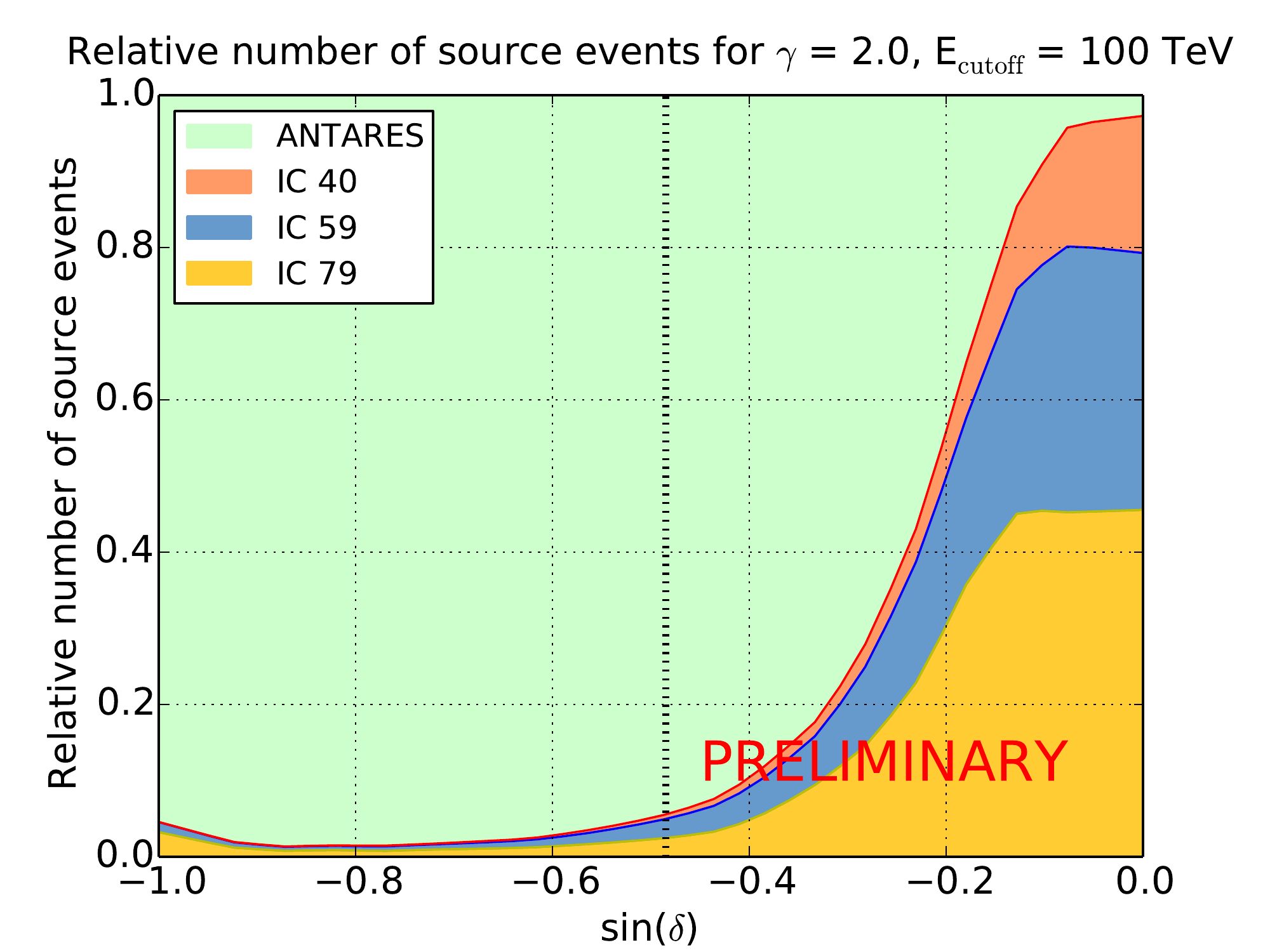}
    \includegraphics[width=.35\textwidth]{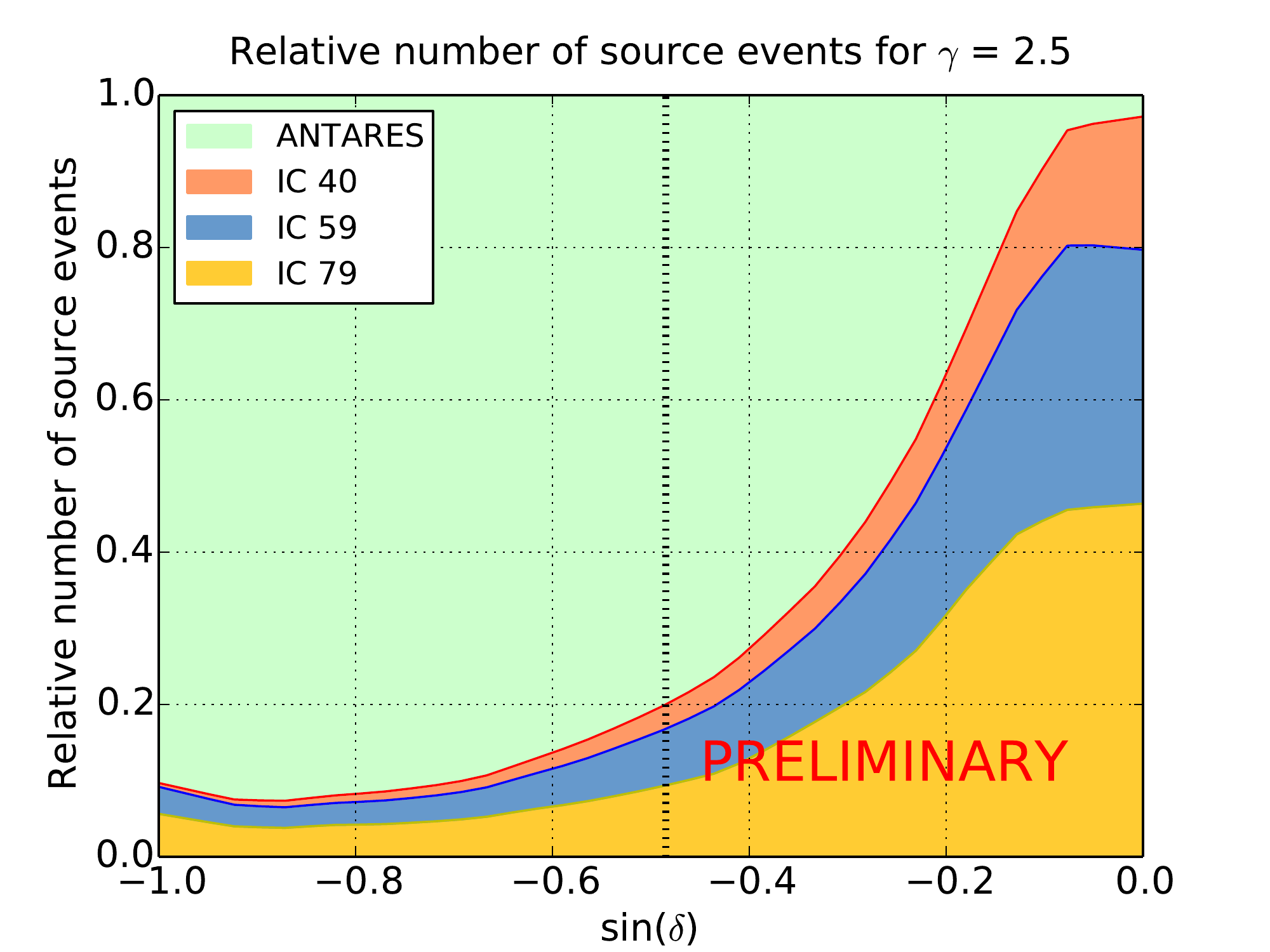}
	\caption{Relative fraction of signal events of each sample as a function of the source declination for different energy spectra: $E^{-2}$ with energy cutoff $E_{\rm cutoff}$ of 1 PeV (top-left), 300 TeV (top-right), 100 TeV (bottom-left); and $E^{-2.5}$ spectrum (bottom-right). The orange, blue and yellow shaded areas correspond to the IceCube 40, 59 and 79-string data samples, respectively, and the green shaded area indicates the ANTARES sample. The relative fraction is used as part of the likelihood function calculation during the search.}
	\label{RelContCases}
\end{figure}

Other source assumptions are also considered in this analysis. The relative fraction of source events is also calculated  for an unbroken $E^{-2.5}$ power-law spectrum, as suggested in recent IceCube diffuse-flux searches \cite{IceCube-Diffuse}, and for an $E^{-2}$ spectrum with an exponential square-root cut-off
 ($\frac{d\Phi}{dE} \propto E^{-2} \exp \left[ { - \sqrt{ \frac{E}{E_{\rm{cut-off}} } } } \right] $)
 for energy cut-offs of 100 TeV, 300 TeV and 1 PeV, since a square-root dependence may be expected from Galactic sources \cite{KAPPES}. 
Figure \ref{RelContCases} shows the relative fraction of source events for these cases. Compared with an unbroken $E^{-2}$ spectrum, the contribution of high energy neutrinos in all of these cases is lower, and therefore the relative contribution of the ANTARES sample increases. 

\section{Search method}\label{method}

An unbinned maximum likelihood ratio estimation has been performed to search for excesses of events that could indicate cosmic neutrinos coming from a source. In order to estimate the significance of a cluster of events, this likelihood takes into account the energy and directional information of each event. The  data sample to which an event belongs is also taken into account, due to the differences in detector response. The likelihood, as a function of the total number of fitted signal events, $n_s$, can be expressed as

\begin{equation}\label{eq:likelihood}
	L(n_s) = \prod_{j=1}^4 \prod_{i=1}^{N^j} \left[ \frac{n^j_s}{N^j}S^j_i + \left(1-\frac{n^j_s}{N^j} \right) B^j_i \right] ,
\end{equation}

\par \noindent where $j$ indicates one of the four data samples (ANTARES, IC40, IC59 or IC79), $i$ indicates an event belonging to the $j$-th sample, S$_i^j$ is the value of the signal probability distribution function (PDF) for the $i$-th event in the $j$-th sample, B$^j_i$ indicates the value of the background PDF, $N^j$ is the total number of events in the $j$-th sample, and $n^j_s$ is the number of signal events fitted for in the $j-$th sample. Since a given evaluation of the likelihood refers to a single source hypothesis at a fixed sky location, the number of signal events $n^j_s$ that is fitted for in each sample is related to the total number of signal events $n_s$ by the relative contribution of each sample, $n^j_s = n_s \cdot C^{j}(\delta, \frac{d{\Phi}}{dE})$.

The signal and background PDFs for the IceCube and ANTARES samples have slightly different definitions. The signal PDF for ANTARES is defined as
\begin{equation}
	S^{ANT} = \frac{1}{2\pi\sigma^2} \exp\left(-\frac{\Delta \Psi(\vec{x}_s)^2}{2\sigma^2}\right) P^{ANT}_{s}(\mathcal{N}^{hits}, \sigma) ,
\end{equation}

\par \noindent where $\vec{x_s}$ = ($\alpha_s$, $\delta_s$) indicates the source direction in equatorial coordinates, $\Delta \Psi(\vec{x}_s)$ is the angular distance of a given event to the source, $\sigma$ is the angular error estimate, and P$_s^{ANT}(\mathcal{N}^{hits}, \sigma)$ is the probability for a signal event to be reconstructed with an angular error estimate of $\sigma$ and a number of hits taken in the event reconstruction $\mathcal{N}^{hits}$. The number of hits is a proxy for the energy of the event \cite{PS-ANTARES-4y}. 

The definition of the signal PDFs for the IceCube samples is similar,

\begin{equation}
	S^{IC} = \frac{1}{2\pi\sigma^2} \exp\left(-\frac{\Delta \psi(\vec{x}_s)^2}{2\sigma^2}\right) P^{IC}_{s}(\mathcal{E}, \sigma|\delta) ,
\end{equation}

\par \noindent where the main difference lies in the use of the reconstructed energy, $\mathcal{E}$, and the declination dependence of the probability for a signal event to be reconstructed with a given $\sigma$ and $\mathcal{E}$. Details about the reconstructed energy proxy can be found in \cite{PS-IceCube-79} and \cite{PS-IceCube-40}. The declination dependence is needed mainly because of the event selection cut on reconstructed energy, which is designed to reduce the atmospheric muon background. 

Background events are expected to be distributed uniformly in right ascension. The background PDFs are in fact obtained from the experimental data itself. The definitions of the PDFs are:

\begin{equation}
	B^{ANT} = \frac{B^{ANT}(\delta)}{2\pi} P^{ANT}_{b}(\mathcal{N}^{hits}, \sigma), \;\;\;
	B^{IC} = \frac{B^{IC}(\delta)}{2\pi} P^{IC}_{b}(\mathcal{E}, \sigma|\delta) ,
\end{equation}

\par \noindent where B($\delta$) is the per-solid-angle rate of observed events as a function of the declination in the corresponding sample. $P^{ANT}_{b}(\mathcal{N}^{hits}, \sigma)$ and $P^{IC}_{b}(\mathcal{E}, \sigma|\delta)$ characterize the distributions for background event properties, in analogy with the definitions of $P^{ANT}_{s}$ and $P^{IC}_{s}$ for signal events given above.

The test statistic, TS, is determined from the likelihood (Eq.~\ref{eq:likelihood}) as TS = $\log L(\hat{n}_s) - \log L(n_s=0)$, where $\hat{n}_s$ is the value that maximizes the likelihood. The larger the TS, the lower the probability (p-value) of the observation to be produced by the expected background. Simulations are performed to obtain the distributions of the TS. The significance (specifically, the p-value) of an observation is determined by the fraction of TS values which are larger or equal to the observed TS.

The TS is calculated as a preliminary step to obtain the post-trial p-values of a search. TS distributions for the fixed-source, background-only hypotheses have been calculated in steps of 1$^\circ$ in declination from pseudo-data sets of randomized data. Because these distributions vary with declination, the preliminary TS is turned into a "pre-trial p-value" by comparing the TS obtained at the source location being examined to the background TS distribution for the corresponding declination.  Post-trial significance is then estimated with pseudo-data sets and according to the type of search, as explained together with the results in Section \ref{results}.

Two different searches for point-like neutrino sources have been performed. In the candidate list search, a possible excess of neutrino events is looked for at the location of 40 pre-selected neutrino source candidates. Since the location of these sources is fixed (at known locations with an uncertainty below the angular resolution of all samples) only the number of signal events $n_s$ is a free parameter in the likelihood maximisation. These candidates correspond to all sources in the Southern sky considered in the previous candidate-source list searches performed in the ANTARES and IceCube point-source analyses \cite{PS-ANTARES} \cite{PS-IceCube-79}.

The second search is a ``full sky'' search, looking for a significant point-like excess anywhere in the Southern sky. For this purpose, the likelihood is evaluated in steps of 1$^\circ \times $~1$^\circ$ over the whole scanned region. Since the angular resolution of both telescopes is smaller than the cell size, the source position is taken as an additional free parameter of the likelihood to fit the best position within the boundaries. 

Both the full Southern sky and candidate-list searches have been performed using an $E^{-2}$
source spectrum in the signal PDFs.  The main virtue of the energy term in the PDFs is to add power to distinguish signal neutrinos from the softer spectra of atmospheric neutrinos ($\sim E^{-3.7}$) and atmospheric muons ($\sim E^{-3}$). Limits for the sources in the candidate list have also been calculated for the source spectra mentioned in section  \ref{sample}.

\section{Results}\label{results}

No significant event clusters are found over the expected background. The most significant cluster in the full-Southern sky search is located at equatorial coordinates $\alpha$ = 332.8$^\circ$, $\delta$=--46.1$^\circ$, with best-fit $\hat{n}_s=7.9$ and pre-trial p-value of $6.0\times10^{-7}$.  
It's found that 24\% of pseudo-data sets have a smaller p-value somewhere in the sky than is found in the real data;  the post-trial significance is thus 24\% (0.7$\sigma$ in the one-sided sigma convention).  The direction of this cluster is consistent with the second most significant cluster in the previous ANTARES point-source analysis (but also less significant). 

\begin{table}[h]
\tiny
\centering
\begin{tabular}{lccccccccc}
\toprule
Name   & $\delta$ ($^\circ$)  &  $\alpha$ ($^\circ$) & $n_s$ &
$p$    & $\phi^{90CL}_{ E^{-2}}$ & $\phi^{90\%CL}_{E_{c} = 1PeV}$  & $\phi^{90CL}_{E_{c} = 300 TeV}$ &
$\phi^{90CL}_{E_{c} = 100 TeV}$ & $\phi^{90CL}_{E^{-2.5}}$ \\
\midrule

HESSJ1741-302   & -30.2         & -94.8         & 1.6   & 0.003         & 2.5E-08       & 7.5E-06       & 5.5E-08       & 7.2E-08       & 1.0E-07\\
3C279   & -5.8  & -166.0        & 1.1   & 0.05  & 3.1E-09       & 1.0E-06       & 6.5E-09       & 9.2E-09       & 6.7E-08\\
PKS0548-322     & -32.3         & 87.7  & 0.9   & 0.07  & 1.6E-08       & 5.0E-06       & 3.8E-08       & 4.9E-08       & 1.4E-08\\
ESO139-G12      & -59.9         & -95.6         & 0.8   & 0.07  & 1.8E-08       & 3.9E-06       & 2.9E-08       & 3.7E-08       & 5.1E-08\\
HESSJ1023-575   & -57.8         & 155.8         & 0.8   & 0.08  & 1.7E-08       & 3.5E-06       & 2.8E-08       & 3.5E-08       & 4.7E-08\\
RCW86   & -62.5         & -139.3        & 0.2   & 0.11  & 1.4E-08       & 4.4E-06       & 3.6E-09       & 4.0E-08       & 5.7E-08\\
\bottomrule

\end{tabular}
\caption{Pre-trial p-values, $p$, fitted number of source events, $n_s$, and 90\% C.L. flux limits, $\Phi^{90CL}_{\nu}$ for the different source spectra for the 6 candidate sources with the lowest p-values. Units for the flux limits for the E$^{-2.5}$ spectra, $\phi^{90CL}_{E^{-2.5}}$,  are given in $GeV^{1.5}cm^{-2}s^{-1}$, whereas the rest are in $GeV cm^{-2}s^{-1}$. The sources are sorted by their declination. }
\label{tab:CL}
\end{table}

The results of the candidate source list search are presented in Table\,\ref{tab:CL}.  No statistically significant excess is found.  The most significant excess for any object in the list corresponds to HESS\,J1741-302 with a pre-trial p-value of 0.003.
 To account for trial factors, the search is performed on the same list of sources using pseudo data-sets
 .  11\% of randomized data sets have a smaller p-value for some source than that found for the real data; the post-trial significance of the source list search is thus 11\% ($1.2\sigma$ in the one-sided sigma convention).  

\begin{figure}[!th]
	\centering
	\includegraphics[width=.48\textwidth]{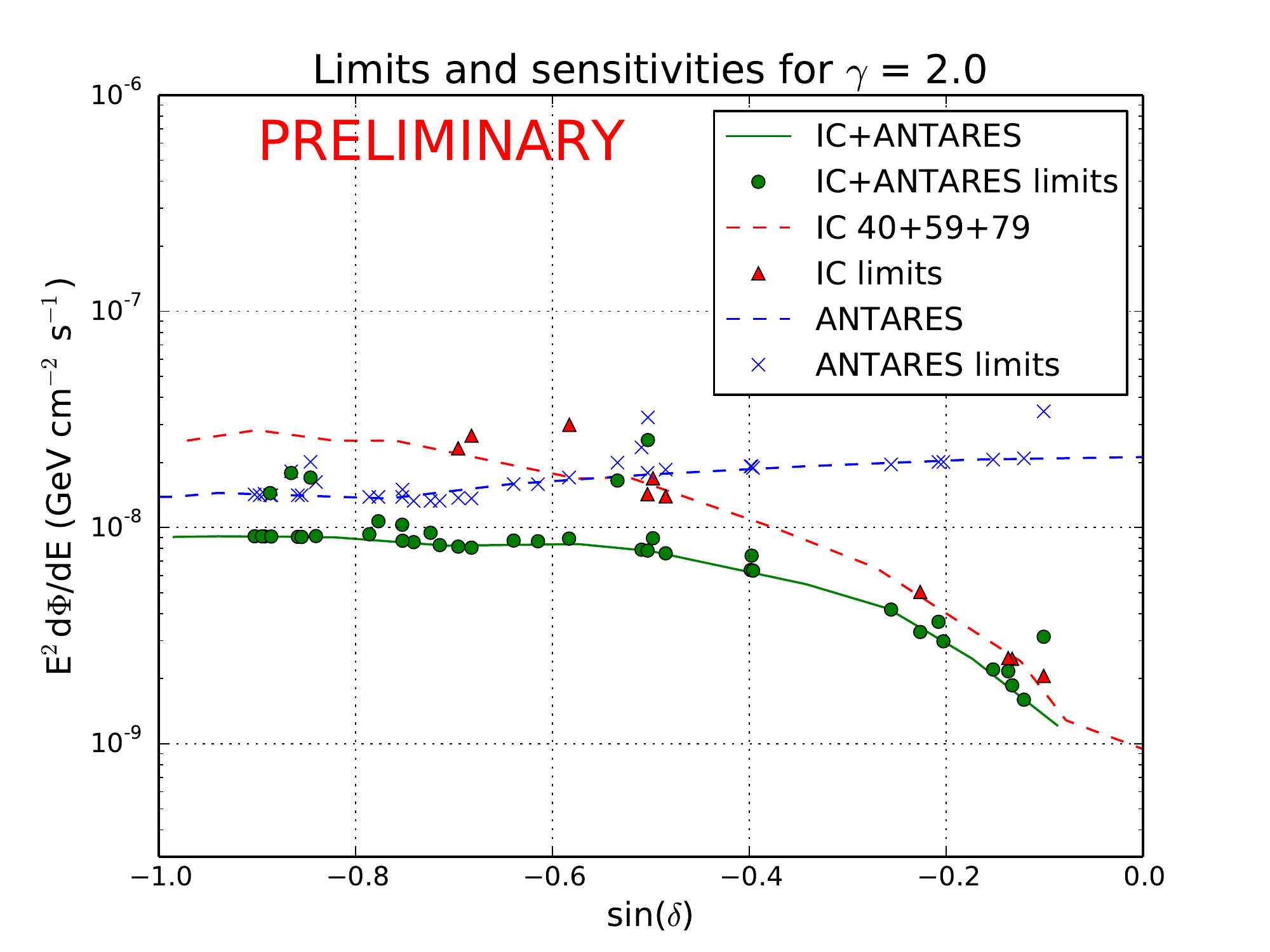}
	\caption{90\% CL sensitivities and limits (Neyman method) for the neutrino emission from point sources as a function of source declination in the sky, for an assumed $E^{-2}$ energy spectrum of the source.  Green points indicate the actual limits on the candidate sources. The green line indicates the sensitivity of the combined search. Blue and red curves/points indicate the published sensitivities/limits for the IceCube and ANTARES analyses, respectively.}
	\label{sensX20}
\end{figure}

\begin{figure}[!th]
	\centering
	\includegraphics[width=.44\textwidth]{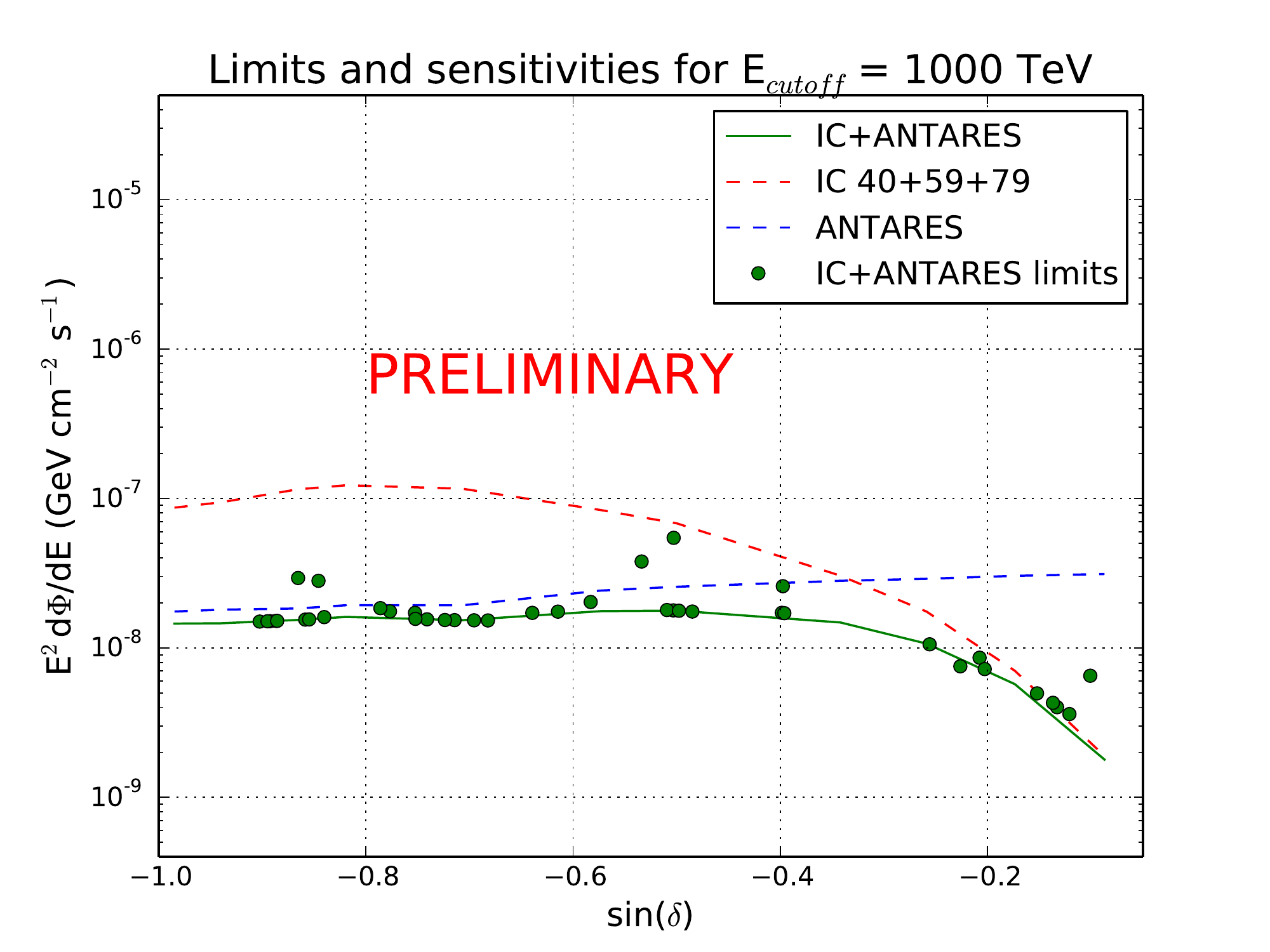}
	\includegraphics[width=.44\textwidth]{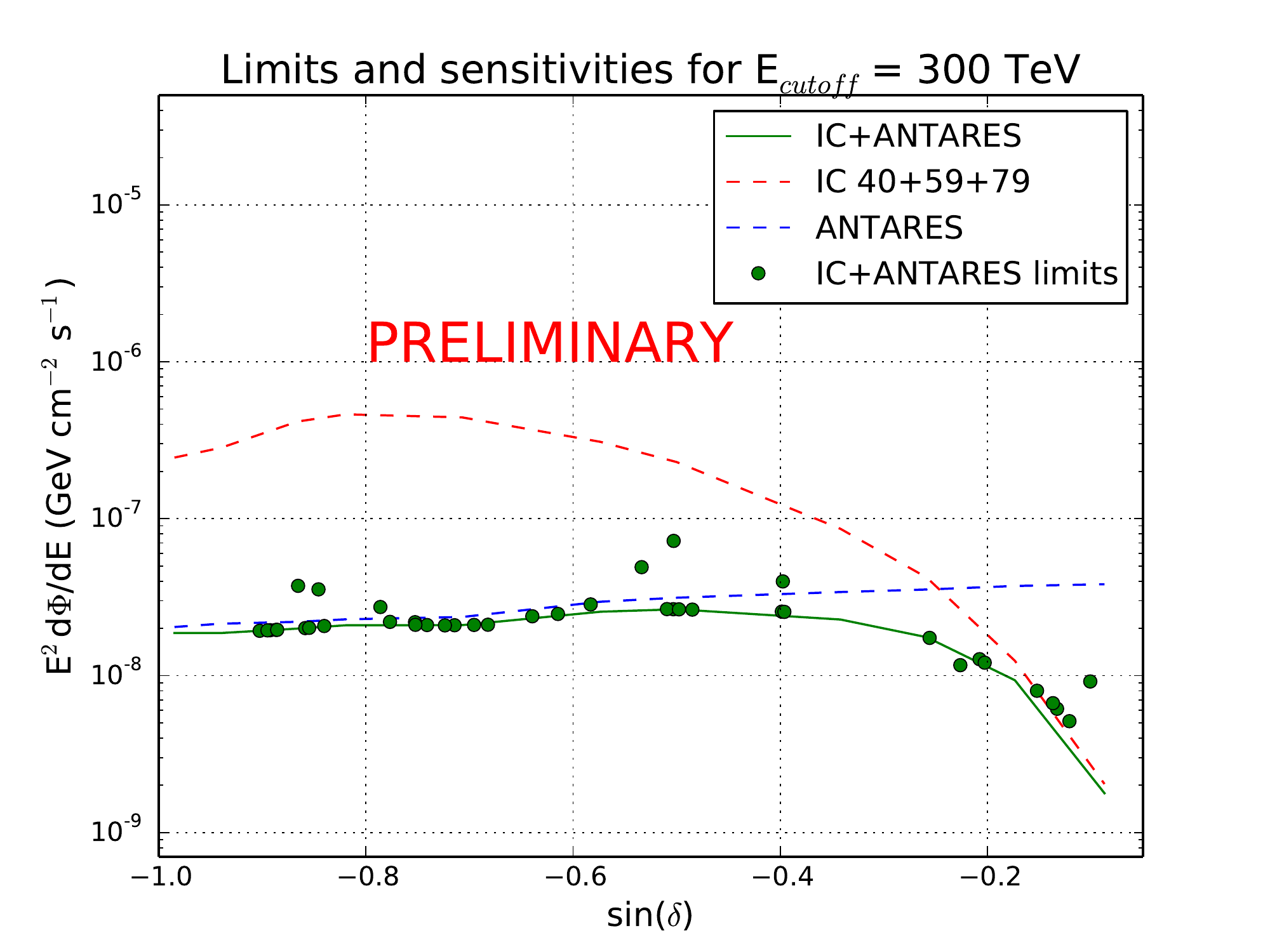}
    \includegraphics[width=.44\textwidth]{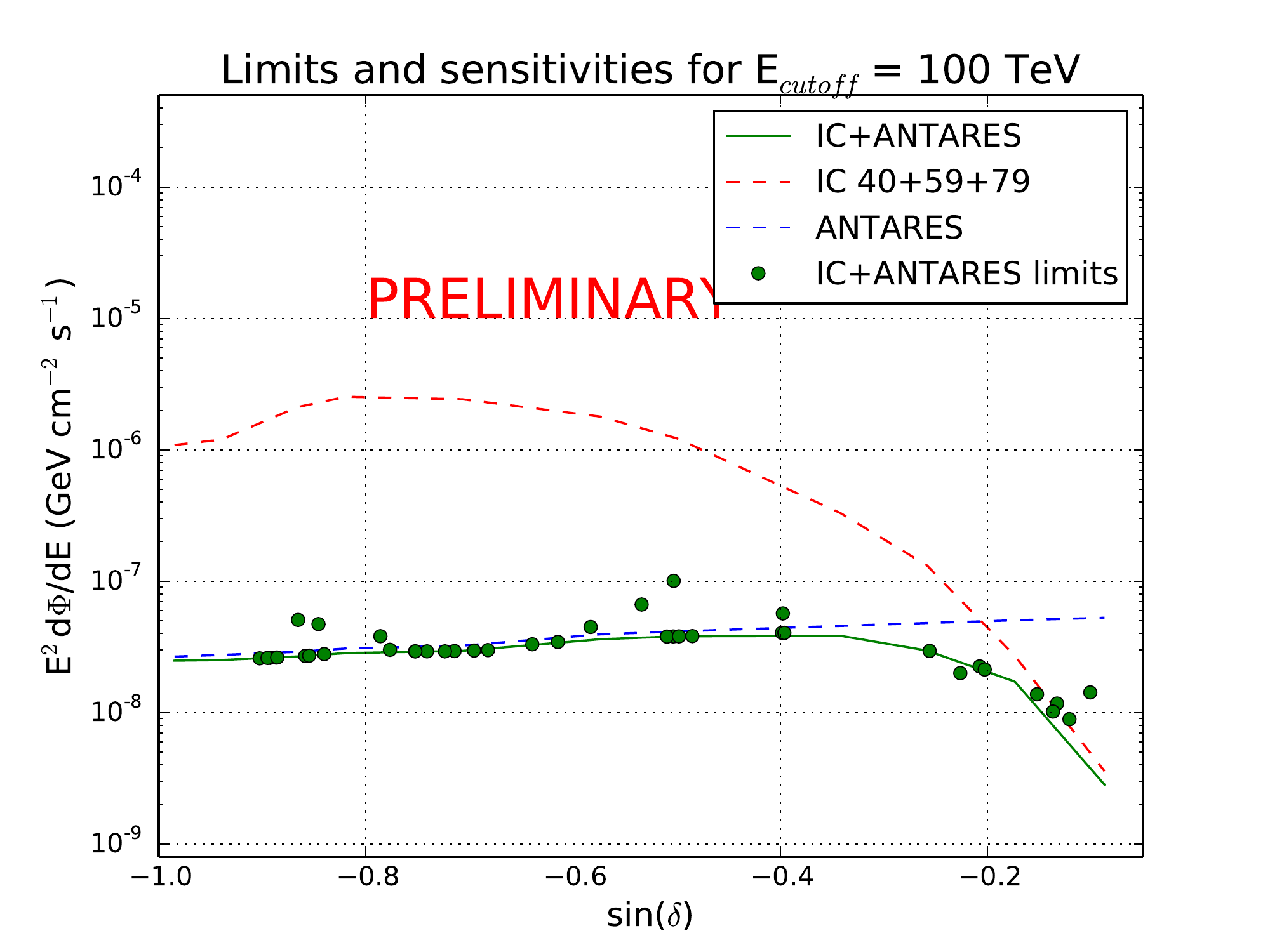}
    \includegraphics[width=.44\textwidth]{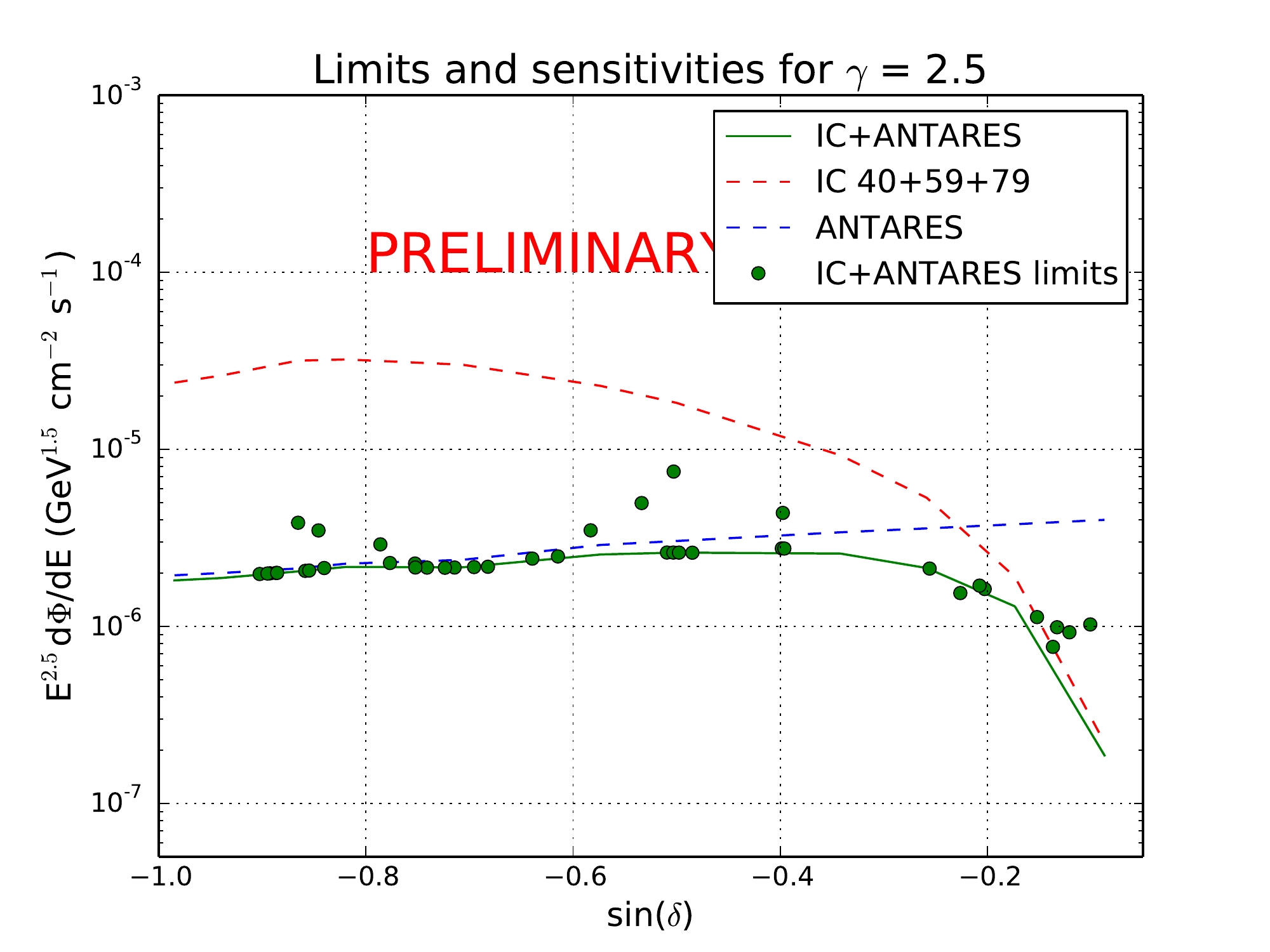}
	\caption{Point source sensitivities and limits for the following energy spectra: $E^{-2}$ with a square-root exponential cut-off at $E = 1\,$PeV (top left), $E = 300\,$TeV (top right), $E = 100\,$TeV (bottom left) and $E^{-2.5}$ unbroken power-law (bottom right). Green points indicate the actual limits on the candidate sources. The green line indicates the sensitivity for the combined search. Blue and red curves/points indicate the sensitivities for the individual IceCube and ANTARES analyses, respectively.  }
	\label{sensCases}
\end{figure}

Table \ref{tab:CL} provides the pre-trial p-values, best-fit signal events $n_s$ and flux upper limits (under different assumptions of the energy spectrum) for the six sources with the lowest p-value.  Figure~\ref{sensX20} shows the sensitivities and limits for this search (assuming an $E^{-2}$ spectrum) in comparison with the previously published ANTARES and IceCube analyses of the same data.  The point source sensitivity in a substantial region of the sky, centered approximately at the declination of the Galactic Center ($\delta=-30^{\circ}$), can be seen to have improved by up to a factor of two.  Similar gains in other regions of the sky can be seen for different energy spectra in Figure \ref{sensCases}.

\section{Conclusion}\label{conclusions}

We have presented the first combined point-source analysis of data from the ANTARES and IceCube detectors.  The combination of their different characteristics, in particular IceCube's larger size  and ANTARES' location in the Northern hemisphere, complement each other for Southern sky searches. We have calculated the sensitivity to point sources and, with respect to an analysis of either data set alone, found that up to a factor of two improvement is achieved in different regions of the Southern sky, depending on the energy spectrum of the source.  Two joint analyses of the data sets have been performed: a search over the whole Southern sky for a point-like excess of neutrino events, and a targeted analysis of 40 pre-selected candidate source objects.  The largest excess in the Southern sky search has a post trial p-value of 0.24 (significance of 0.7$\sigma$). In the source list search the candidate with the highest significance corresponds to HESS\,J1741-302, with a post-trial p-value of 0.11 (significance of 1.2$\sigma$). Both of the results are compatible with the background-only hypothesis and no significant excess is found. Flux upper limits for each of the source candidates have been calculated for $E^{-2}$ and $E^{-2.5}$ power-law energy spectra, as well as for $E^{-2}$ spectra with cut-offs at energies of 1\,PeV, 300\,TeV, and 100\,TeV.  Because of their complementary nature, with IceCube providing more sensitivity at higher energies and ANTARES at lower energies, a joint analysis of future data sets will continue to provide the best point-source sensitivity in critical overlap regions in the Southern sky, where neutrino emission from Galactic sources in particular may be found.

%\acknowledgments
% Neutrins, vos estimem! :) (Tot i que vosaltres paseu de nosaltres en la grandísima majoria de casos...)

%\includepdf[pages={1-},scale=1,offset=72 -72]{Combined.pdf}

\end{document}